\documentclass[reprint, showpacs, preprintnumbers, amsmath, amssymb, aps, prb,url,]{revtex4-1}
\usepackage{color,soul}
\usepackage{graphicx}
\usepackage{bm}
\usepackage{dcolumn}
\usepackage{sidecap}
\usepackage{amsmath}
\usepackage[version=4]{mhchem}

\begin{document}

\bibliographystyle{apsrev4-1}

\title{Magnetite nano-islands on silicon-carbide with graphene}

\author{Nathaniel A. Anderson$^{a}$, Qiang Zhang$^{a}$, Myron Hupalo$^{a}$, Richard A. Rosenberg$^{b}$, Michael C. Tringides$^{a}$, and David Vaknin$^{a}$}
\affiliation{
$^a$Ames Laboratory and Department of Physics and Astronomy, Iowa State University, Ames, Iowa 50011, United States\\
$^b$X-ray Science Division, Advanced Photon Source, Argonne National Laboratory, Lemont, Illinois 60439, United States\\
}

\date{\today}

\begin{abstract}
 X-ray magnetic circular dichroism (XMCD) measurements of iron nano-islands grown on graphene and covered with a Au film for passivation reveal that the oxidation through defects in the Au film spontaneously leads to the formation of magnetite nano-particles (i.e, \ce{Fe3O4}). The Fe nano-islands   (20 and 75 monolayers; MLs) are grown on epitaxial graphene formed by thermally annealing 6H-SiC(0001) and subsequently covered,  in the growth chamber, with nominal 20 layers of Au. Our X-ray absorption spectroscopy and XMCD measurements at applied magnetic fields show that the thin film (20 ML) is totally converted to magnetite whereas the thicker film (75 ML) exhibits properties of magnetite but also those of pure metallic iron. Temperature dependence of the XMCD signal (of both samples) shows a clear transition at $T_{\rm V}\approx 120$ K consistent with the Verwey transition of bulk magnetite.  These results have implications on the synthesis of magnetite nano-crystals and also on their regular arrangements on functional substrates such as graphene.
\end{abstract}
\maketitle

\section*{Introduction}
Creating magnetic nano-crystals (NCs), and in particular magnetic oxides, is by now a common practice; however, organizing them in a regular structure on a functional substrate, such as graphene, is still a challenge.  Magnetite, the naturally-occurring magnet, and its derivatives have been produced in NC forms by a few methods that include co-precipitation, thermal decomposition and/or reduction, micelle synthesis, hydrothermal synthesis, laser pyrolysis, and others\cite{Lu2007,Wu2015,Xu2014}. In nature, NCs magnetites (100-200 nm in size) are found to be embedded in lodestone surrounded by \ce{Fe2TiO4}\cite{Harrison2002}. Another intriguing natural occurrence of magnetite NCs (50 - 100 nm in size) is in magneto-tactic bacteria that utilize them for orientation with respect to geomagnetic fields presumably for navigation to oxygen rich aqueous regions\cite{Blakemore1975}. Recently, a membrane protein that promotes the growth of magnetite has been used to mimic this  biomineralization process in the laboratory\cite{Wang2012}.  Assembling  other transition-metal oxides on graphene has gained impetus in recent years\cite{Wang2010}. Self-assembled graphene/\ce{Fe3O4} hydrogels with robust interconnected 3D networks have been fabricated on a large scale  induced by  Fe(II) ions at different pH values,\cite{Zhou2010} and a related study describes the fabrication of graphene nanosheets that are decorated with \ce{Fe3O4} particles by in-situ reduction of iron hydroxide to form anode materials for  Li batteries\cite{Cong2012}. Graphene is chosen as a substrate for Fe NCs  growth because of its potential applications in microelectronics, catalysis, and spintronics\cite{Hershberger2013,Dankert2014}.  Here, we report on the chemical and magnetic properties of pure iron-metal nano-islands, discretely  arranged on graphene and capped with a gold film for protection against oxidation when removed from the growth chamber and  transferred in air for subsequent experiments.  We employ X-ray absorption spectroscopy (XAS) and X-ray magnetic circular dichroism (XMCD) to determine specifically the chemical species and their magnetic properties.  

\section*{Experimental Details}
\begin{figure}\centering \includegraphics [width = 3 in] {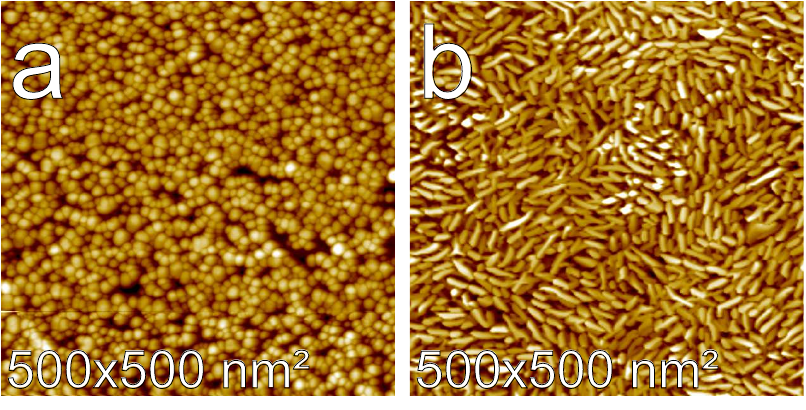}
\caption{STM images of  (a) 20 ML (thin) and (b) 75 ML (thick) pure Fe metal deposited on graphene. Samples were capped with 20 ML Au in the high vacuum chamber for passivation against oxidation so that they can be transported (in-air) for the  XAS and XMCD measurements.}
\label{fig:FeSTM} 
\end{figure}

Samples are prepared by depositing Fe on graphene grown on a SiC substrate\cite{Hershberger2013} using a molecular beam source held at a temperature of 700 K with a flux rate of 0.1 - 0.2  monolayers (ML)/min. The Fe source is degassed during the bake-out for several hours, so that during deposition the pressure remains below $1.6\cdot10^{-10}$ Torr.  Figure \ref{fig:FeSTM} shows STM images of the (a) thin (20 ML) and (b) thick (75 ML) samples of Fe on graphene\cite{Hershberger2013}.  As shown in the STM images, islands of approximately 10 nm diameter are formed in the shape of pillars of roughly uniform height. The average number of Fe-MLs in each island is determined by evaluating each island's volume within a given area after correcting for finite size tip effects. After the STM imaging and while still under ultra-high vacuum, both samples were capped with nominal 20 layers of Au for protection against oxidation during sample transport (in air) for the XMCD experiments.
XMCD measurements were performed at the  4-ID-C beamline at the Advanced Photon Source (Argonne National Laboratory) in a chamber equipped with a high magnetic field ($<6$ T) produced by a split-coil superconducting magnet. Field dependence of the XMCD spectra were collected in helicity-switching mode in external magnetic fields applied parallel to the incident x-ray wave vector at energies that covered the Fe $L_2$  (719.9 eV) and $L_3$  (706.8 eV)  binding energies.   The X-ray incident angle was fixed at $\approx 12 \pm 2$ degrees with respect to the sample surface and measurements of x-ray absorption spectroscopy (XAS) signals were collected by  total electron yield (TEY), reflectivity (REF), and fluorescence. REF is detected with a Si diode and TEY is determined from the drain current to the sample. All detection modes are collected simultaneously.  Here, we report TEY results only, however results in other modes were examined for consistency. 
XMCD signal is obtained from the difference between two XAS spectra of the left- and right-handed helicities, $\mu_+$  and $\mu_-$. All XAS signals values for $\mu_+$  and $\mu_-$ are normalized to the incident beam monitor intensity. For detailed data normalization and background evaluation see SI. 

\begin{figure}\centering \includegraphics [width = 3 in] {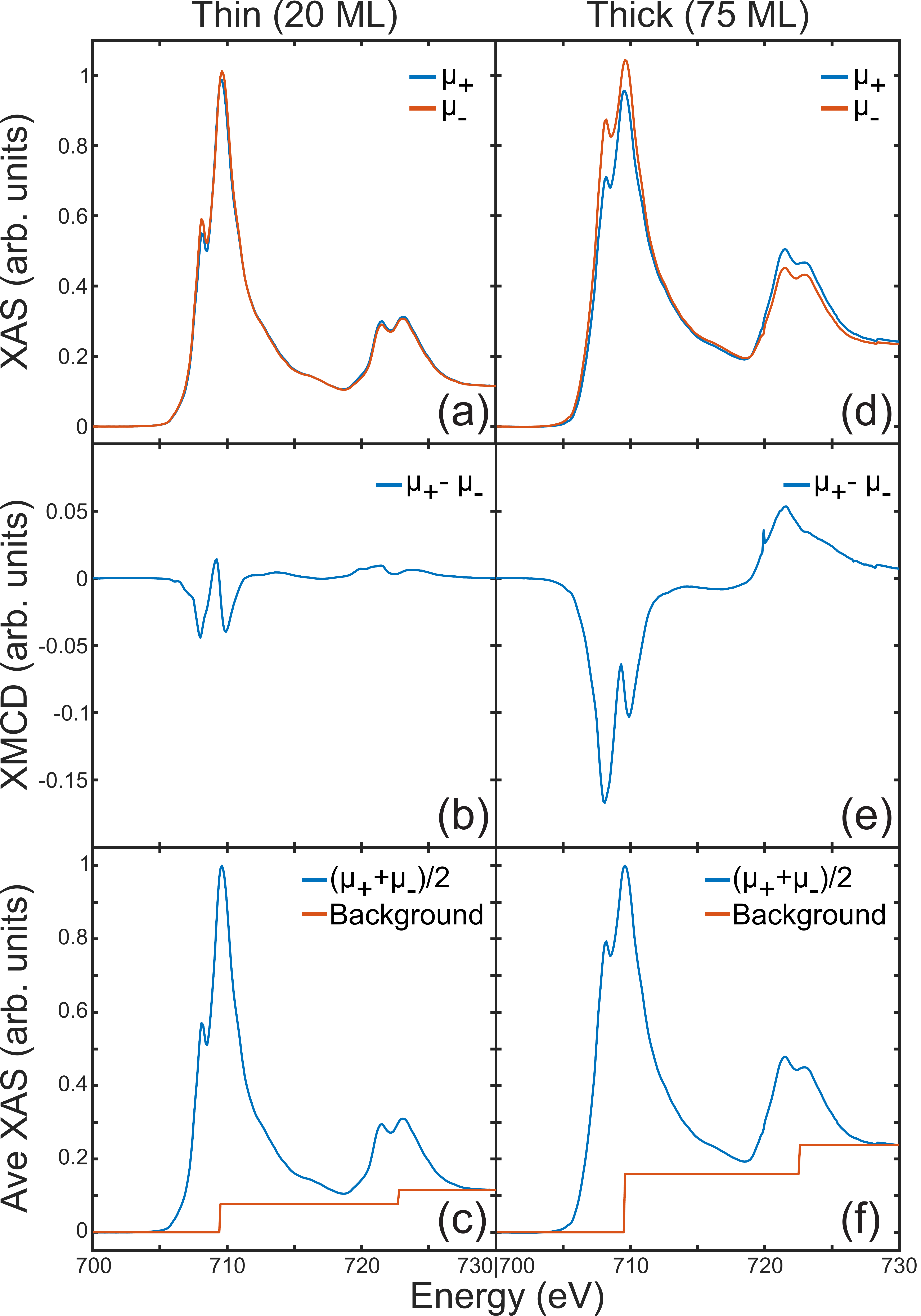}
\caption{(a), (b), and (c) are the normalized XAS, XMCD, and average XAS of the thin Fe sample (20 ML high islands), respectively.  (d), (e), and (f) are the XAS, XMCD, and average XAS of the thick Fe sample (75 ML high islands), respectively. Both samples are grown on graphene at B = 5 T and T = 15 K.}
\label{fig:Thin-XAS-XMCD}
\end{figure}

\section*{Results and Discussion}
\begin{figure}\centering \includegraphics  [width = 3 in] {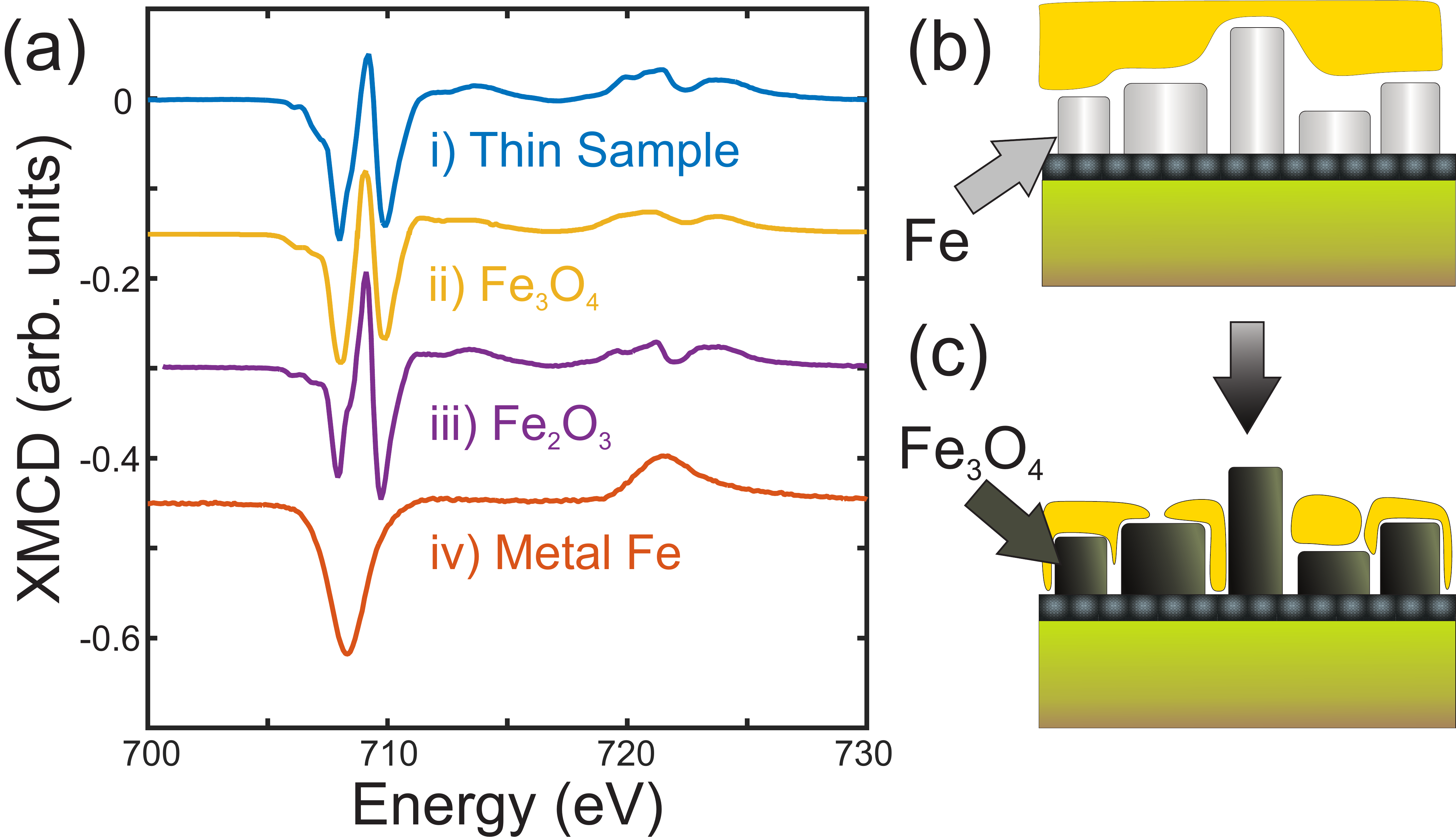}
\caption{ (a) The XMCD spectra for (i) thin Fe nano-islands sample at $B=5$ T and $T=15$ K compared to those of (ii) \ce{Fe3O4}, (iii) \ce{Fe2O3}, and (iv) metal Fe. (b) Depiction of tight gold film on iron islands intended to prevent oxidation  (c) Depiction of defects in the gold film that allow oxidation of Fe nano-islands to form nearly pure \ce{Fe3O4}.The spectra for $\ce{Fe2O3}$ (iii) and Fe (iv) were graciously provided by the team at Sector 4 of the Advanced Photon Source and the $\ce{Fe3O4}$ spectra (ii) was taken from Goering et al.$\cite{Goering2007}$  }
\label{fig:Thin-Comp}
\end{figure}

Figure \ref{fig:Thin-XAS-XMCD} shows normalized XAS and XMCD for both samples (20 and 75 MLs)  at $B = 5$ T and $T= 15$ K over the range of $700-730$ eV. The XAS in the two polarization modes show splittings of the $L_2$ and $L_3$ peaks that are not  characteristic of iron metal suggesting the presence of Fe$^{2+}$ and Fe$^{3+}$ in the sample.  The XMCD spectra for both samples extracted from these two XAS scans (shown in Fig. \ref{fig:Thin-XAS-XMCD}(b and e)) are  significantly different from that reported for iron metal.\cite{Chen1995}  Thus, despite being covered with gold, our samples show strong features that indicate they are possibly oxidized raising the  question of what iron oxide species are formed.  It has been shown that both the XAS  and the XMCD exhibit distinct features that can distinguish among various iron oxides including  FeO, $\alpha -$ and $\gamma$-\ce{Fe2O3}, Fe$_{3-\delta}$O${_4}$, and \ce{Fe3O4}.\cite{Pellegrain1999,Brice-Profeta2005,Huang2004,Martin-Garcia2015,Chen1995} Furthermore, it has been shown that the XMCD spectra of \ce{Fe3O4} is a superposition of three components corresponding to the three distinct sites of iron in this inverse spinel.\cite{Pattrick2002}      In Fig. $\ref{fig:Thin-Comp}$ we compile the XMCD spectra from various iron oxides and iron metal to compare them with that of our thin sample.  We note that the two closest spectra to ours, are those of \ce{Fe3O4} and $\gamma$-\ce{Fe2O3}.  Closer inspection of the two spectra shows that the first two minima (at the $L_3$ edge) are reversed in magnitude between the two, indicating that our sample is closer in composition to \ce{Fe3O4} than to $\gamma$-\ce{Fe2O3}. We therefore conclude that the protection of the nano-islands with 20 MLs of gold does not prevent iron from oxidation, most likely due to defects and incomplete uniform protective layer (see depiction of expected and experimental results in Fig.\ \ref{fig:Thin-Comp} (b) and (c), respectively). More importantly the Fe nano-islands at this size (20 ML thick) readily convert to a majority magnetite \ce{Fe3O4} phase with a possible minority phase of maghemite, $\gamma$-\ce{Fe2O3} (further experimental evidence for the formation of \ce{Fe3O4} is provided below.) This result, demonstrating  that at the nano-scale size (approximately 10 nm in diameter) pure iron if exposed to oxygen transforms spontaneously to magnetite, is  extremely important in view of the fact that tremendous efforts have been dedicated to synthesizing magnetite nano-particles\cite{Lu2007,Wu2015,Xu2014}. This also opens an avenue for assembling magnetite nanoparticles on solid surfaces by first arranging, by deposition, pure iron nano-particles and subsequently exposing them to an oxidative environment. 

\begin{figure}\centering \includegraphics [width = 3 in] {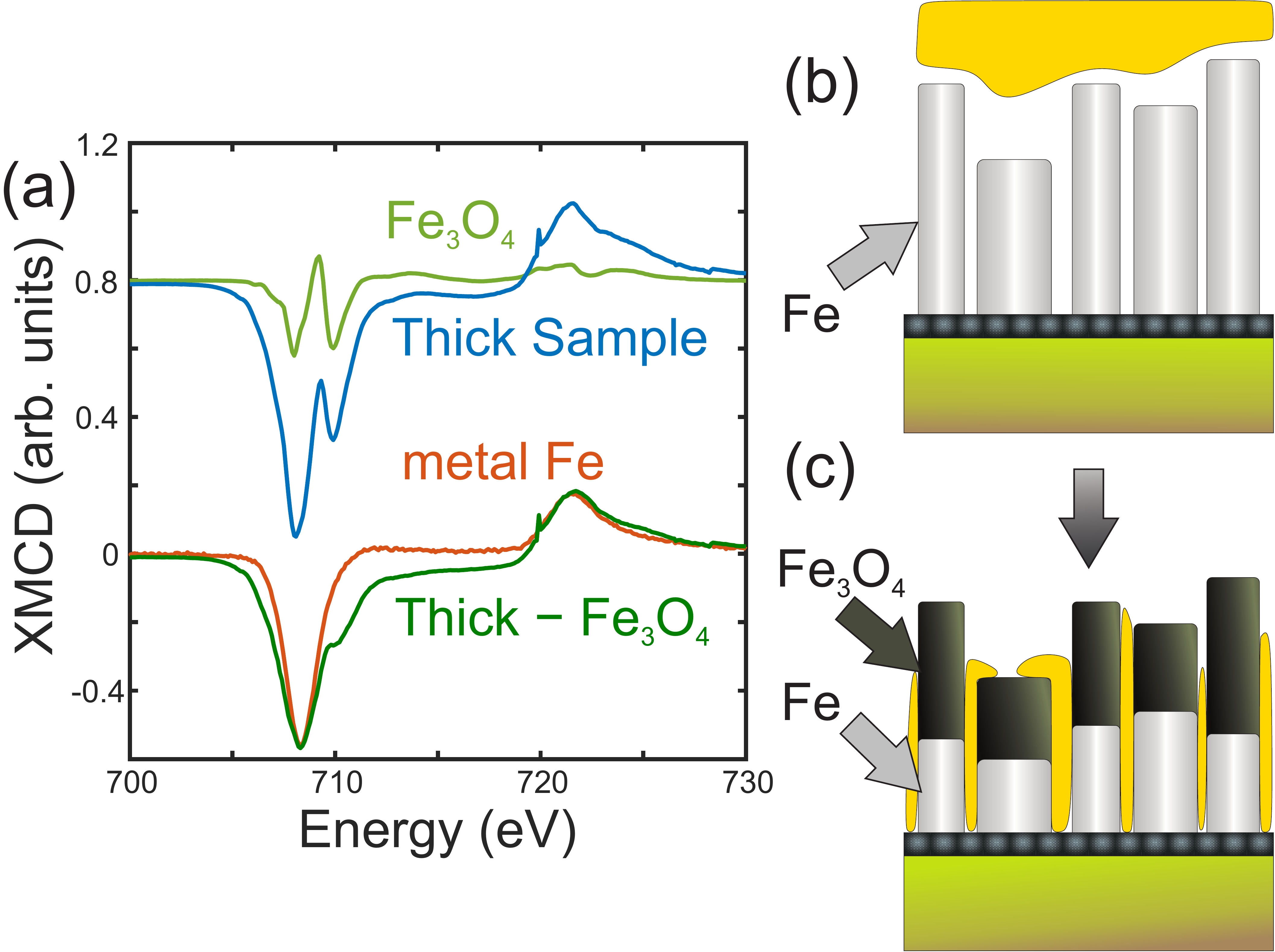}
\caption{(a) The top two spectra show the thick Fe-nano-islands sample on graphene at $B=5$ T and $T=15$ K compared to the thin sample (Fig. \ref{fig:Thin-XAS-XMCD}b) scaled by a factor of 1.4 (assumed to be entirely \ce{Fe3O4}). The bottom two spectra show a comparison of pure metal iron to a subtraction of the top spectra. b) Depiction of tight gold film on iron islands intended to prevent oxidation  (c) Depiction of defects in the gold film that allow oxidation  of Fe nano-islands to partially form of \ce{Fe3O4}. }
\label{fig:Thick-Comp} 
\end{figure}

The XMCD for the thick sample (75 MLs; Fig. \ref{fig:Thick-Comp}) matches neither that of pure Fe nor that of \ce{Fe3O4}. Due to fast electronic relaxation times, the XMCD features from pure iron (or other good metals) are not split and are slightly broadened over those in insulators and semimetals (i.e., magnetite).\cite{Chen1995} Thus, based on the features and linewidths of peaks observed in the XMCD measurements, we suggest that the islands of the thick sample ($\sim 75$ MLs) consist of metallic iron and \ce{Fe3O4}. By applying a scale factor of 1.4 to the XMCD signal from the thin sample (Fig. \ref{fig:Thin-XAS-XMCD}b) and subtracting it from that of the thick sample (Fig. \ref{fig:Thick-Comp}) we obtain a curve that resembles that of metallic iron (Fig. \ref{fig:Thick-Comp}) albeit slightly wider.  We therefore propose that the taller islands are only partially oxidized to form \ce{Fe3O4} and that metal iron is still present in some parts of the sample (as depicted in Fig. \ref{fig:Thick-Comp} (b) and (c)).  The different observations between tall and short islands suggests that the kinetics of oxidation is time dependent (both samples, after being capped by gold, have been in air for about a few weeks outside the ultra-high vacuum chamber). We have, in fact, reexamined  the thick sample after seventeen months and found that its XMCD pattern indicates more conversion into magnetite  (see Supporting Information).  We note that our results on the transformation of pure metal iron to magnetite through defects in a protecting layer are consistent with a recent study that examines the long term efficiency of passivating various transition metal surfaces with a layer of graphene \cite{Weatherup2015}.  That study shows that for a Ni surface, the graphene protects the surface from oxidation and furthermore, at undesired defects in the graphene layer, the oxidation of Ni is localized and prevents  oxidation diffusion into the surface.  On the other hand, for iron surfaces it is found that oxidation at defect points is no longer localized and oxidation practically diffuses throughout the surface over long period of time. Whereas that study confirms oxidation of Fe surfaces via graphene defects, it does not provide chemical analysis of the species that are formed at the surface as provided in this study. 

\begin{figure}\centering \includegraphics [width = 3 in] {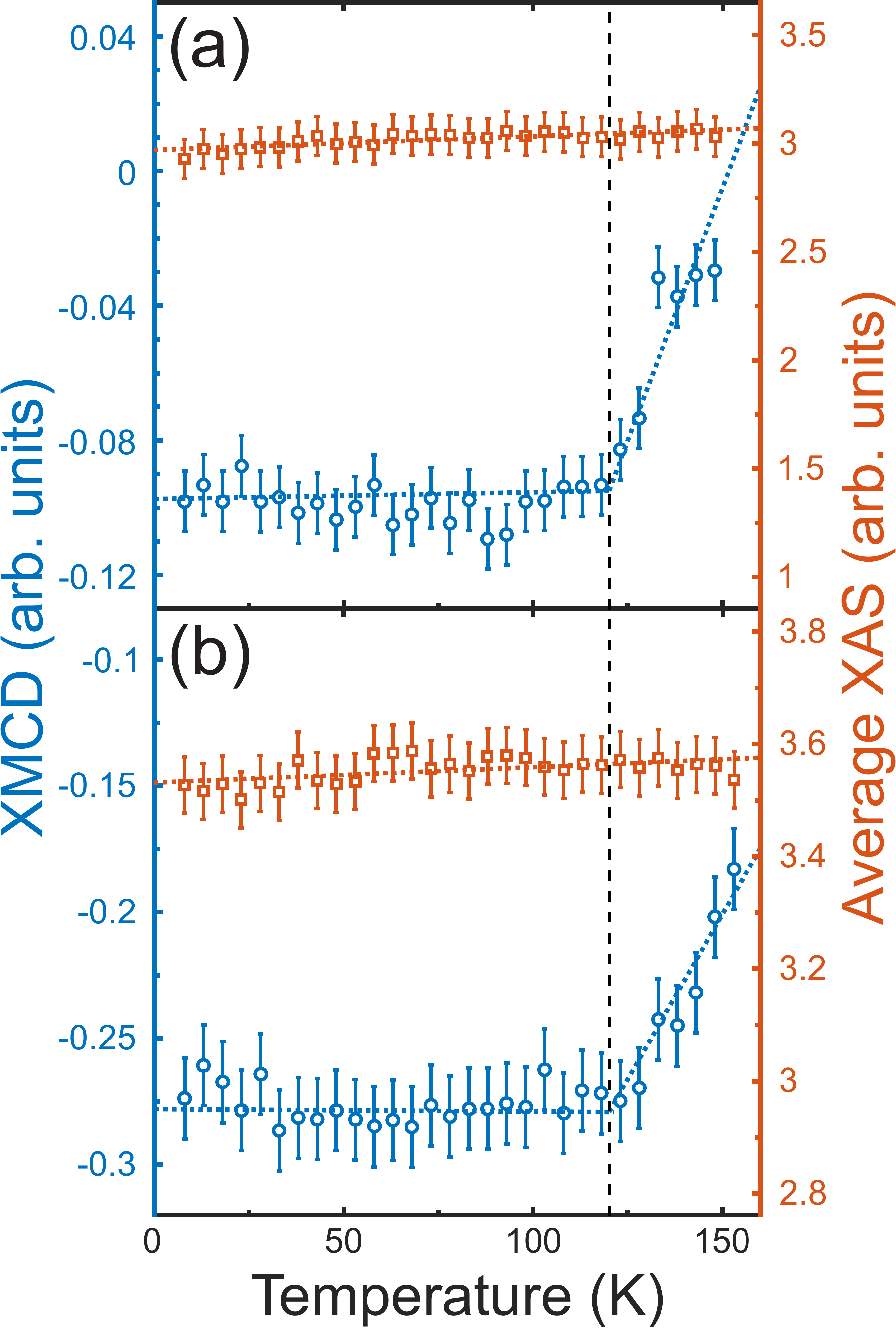}
\caption{The temperature dependence of the XMCD (blue $\circ$ left), average XAS (red $\square$ right) of the integrated intensity at the $L_3$ peak for (a) thin (20 ML) and (b) thick (75 ML) Fe on graphene at $B=3$ T. Vertical dashed lines indicate the Verwey transition at $T_V = 120$ K,  dashed lines through the data points are guides to to the eye to emphasize the  Verwey transition.}
\label{fig:TempDep} 
\end{figure}
\begin{figure}\centering \includegraphics [width = 3 in] {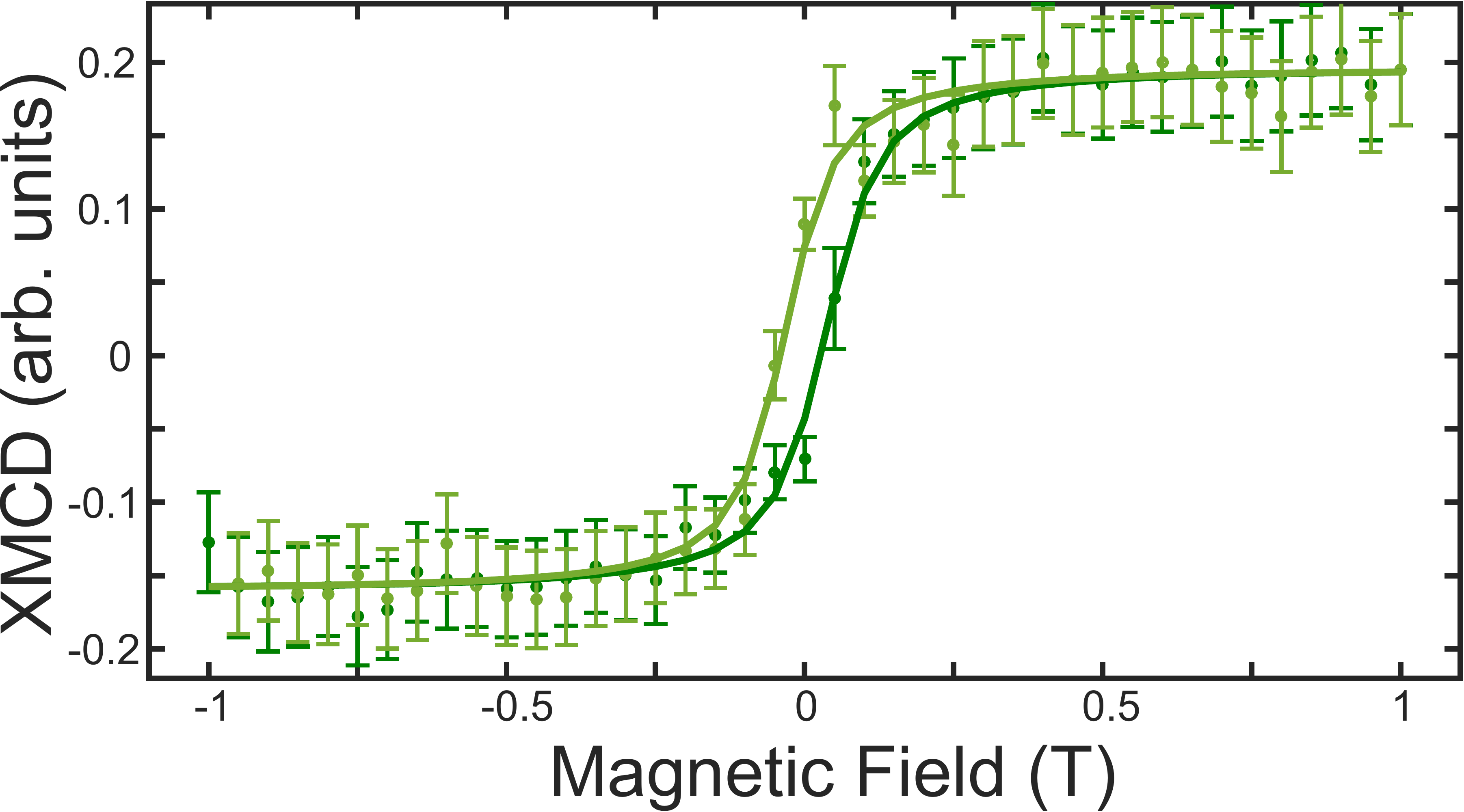}
\caption{Magnetic field dependence of the integrated XMCD signal of the thick sample at $T=15$ K.  The solid lines through measured points are guides to the eye showing the hysteresis curve.}
\label{fig:Hyst} 
\end{figure}

To corroborate our claim on the spontaneous formation of magnetite from pure iron nano-islands by oxidation, we have also systematically looked for signatures typical of magnetite in the temperature dependence of the XAS and XMCD signals from 8 to 153 K. Figure \ref{fig:TempDep} shows the integration of the $L_3$ signal (705-715 eV at $B = 3$ T) of the XMCD and XAS versus temperature with a distinct anomaly at $\sim$120 K for both samples. This anomaly is consistent with the well documented Verwey transition in magnetite which takes place in \ce{Fe3O4} at $\sim$125 K  when undergoing a metal-insulator transition \cite{Senn2012,Yu2014}. The existence of the Verwey transition in nano-size magnetite has been the subject of some debate, however, recently it has been demonstrated that only for particles below $\sim 6$  nm in size is the transition reduced from its bulk value.\cite{Park2005,Lee2015} The change in the integrated $L_3$ XMCD signal  at $T_{\rm V}$ results from the change in magnetic ordered moments and orbital ordering, both of which (spin and orbital moments) contribute to the XMCD.  We note that whereas both samples show the Verwey transition in the  integrated XMCD signals, neither show any change in their XAS signal. The field dependence of the XMCD signal (thick sample) at low temperatures (Fig. \ref{fig:Hyst}), which to a good approximation is proportional to the magnetization of the sample, shows a hysteresis loop with coercivity field $H_c \approx 0.05$  T.   It is well established that for high purity iron  (99.95\%)  $H_c \sim 5\times10^{-6}$T\cite{Brown1958}  thus, the observed hysteresis  is consistent with that of magnetite.  This relatively high $H_c$ value is consistent with similar hysteresis measurements of magnetite that are acicular in shape (needle-like)\cite{Morrish1958,Okuda1985} with similarity to our pillar-like island grown NCs.

\section*{Conclusions}
Using X-ray magnetic circular dichroism (XMCD) of Au-capped Fe nano-islands, we show that iron-oxidation (by exposure to air) proceeds spontaneously, likely through defects in the gold-film. Furthermore, we find that the  iron nano-islands convert to magnetite nano-particles (totally or partially, depending on particle size). The evidence for the transformation of iron nano-islands to magnetite by oxidation is unequivocal as it is also corroborated by the presence of the well  known Verwey transition at $T_V\approx120 $ K as in bulk magnetite.   Our results also indicate that oxidation is size and time dependent. 

\section*{Supplementary Material}
See the Supplementary Material for more information on the following:  (1) The long-term oxidation time dependence of the thick sample's XMCD. (2) Selected XMCD  spectra above and below the Verwey transition, also used for creating the detailed temperature dependence  shown in Figure\ \ref{fig:TempDep} (3) Details on data normalization and background subtraction.

\section{Acknowledgments}
Ames Laboratory is operated by Iowa State University by support from the U.S. Department of Energy, Office of Basic Energy Sciences, under Contract No. DE-AC02-07CH11358. Use of the Advanced Photon Source, an Office of Science User Facility operated for the U.S. Department of Energy (DOE) Office of Science by Argonne National Laboratory, is supported by the U.S. DOE under Contract No. DE-AC02-06CH11357.

\bibliography{FeNanoIslands-JAP2.bbl}
\end{document}